\documentclass[notoc]{JHEP3}
\usepackage[centertags]{amsmath}
\usepackage{amsfonts}
\usepackage{amssymb}
\usepackage{amsthm}
\usepackage{newlfont}
\usepackage{graphicx}
\usepackage{feynmp}

\newcommand{\eps}{\varepsilon}

\newcommand{\Zf}{\mathbb{Z}}
\newcommand{\cn}{\mathcal N}

\newcommand{\ave}[1]{\left\langle #1\right\rangle}

\newcommand{\pdf}[2]{\frac{\partial #1}{\partial #2}}
\newcommand{\diff}[2]{\frac{d #1}{d #2}}

\newcommand{\lag}{\mathcal{L}}
\newcommand{\ra}{\rightarrow}

\newcommand{\slh}[1]{\not\mspace{-4mu}#1}
\newcommand{\slhs}[2]{\not\mspace{-#2mu}#1}
 \DeclareMathOperator{\tr}{Tr}
\newcommand{\prt}{\partial}
\newcommand{\hsp}[1]{\hspace{#1cm}}
\newcommand{\rar}[1]{\hspace{#1em}\Rightarrow\hspace{#1em}}
\DeclareMathOperator{\diag}{diag}

\def\d{\delta}
\def\c{\gamma}
\newcommand{\vef}{V_{\text{eff}}}
\newcommand{\gym}{g_{\text{YM}}}
\def\N{\cal{N}}
\def\O{\cal{O}}
\def\ads{AdS$_5\times$S$^5$ }
\def\nez{$\N=0^*$ }
\def\neo{$\N=1^*$ }
\def\nef{$\N=4$ }
\def\l{\lambda}
\DeclareMathOperator{\SO}{SO}
\title{The Effective Potential of the \nez Yang-Mills Theory}
\preprint{\hepth{0402215}\\WIS/06/04-FEB-DPP}

\author{Assaf Patir and Dori Reichmann\\
Weizmann Institute of Science,\\ Rehovot 76100, Israel\\ E-mail:
\email{assaf.patir@weizmann.ac.il},\email{dor.reichmann@weizmann.ac.il}}

\abstract{We study the $\cn=4$ SYM theory with SU($N$) gauge group
in the large $N$ limit, deformed by giving equal mass to the four
adjoint fermions. With this modification, a potential is
dynamically generated for the six scalars in the theory, $\phi^i$.
We show that the resulting theory is stable (perturbatively in the
't Hooft coupling), and that there are some indications that
$\ave\phi=0$ is the vacuum of the theory. Using the AdS/CFT
correspondence, we compare the results to the corresponding
supergravity computation, i.e. brane probing a deformed
AdS$_5\times S^5$ background, and we find qualitative agreement.}
\begin{document}
\maketitle

\section{Introduction}
The AdS/CFT correspondence relates String Theory on $d+1$
dimensional anti-de Sitter (AdS) space (times some compact
manifold) to a $d$-dimensional conformal field theory (CFT) (see
\cite{Maldacena:1998re,Witten:1998qj,Gubser:1998bc} and the review
\cite{Aharony:1999ti}). In particular, it is conjectured that Type
IIB String Theory on AdS$_5\times S^5$ is equivalent to $\cn=4$
supersymmetric Yang-Mills (SYM) theory with gauge group SU($N$).
In the regime where $\gym^2N$ is large, supergravity is a good
approximation to the full string theory. In this regime, the
identification of partition functions in the two theories allows
one to construct supergravity solutions that are dual to
deformations of the $\cn=4$ SYM theory.

Perhaps the main motivation for the study of such deformed
theories is the prospect of finding a string theory dual to a
field theory that is similar to QCD. This will allow the study of
strong coupling phenomena of QCD-like theories via the
understanding of weakly coupled supergravity (three possible paths
to this goal were reviewed in \cite{Aharony:2002up}). Other
motivations for studying such deformations are learning about
strong coupling phenomena on both sides of the correspondence (by
studying the weakly coupled dual perturbatively) and gaining
insight into how the correspondence works by understanding how
various phenomena are realized in the two dual descriptions.
Eventually, one would hope to go beyond the supergravity
approximation and find the string dual to QCD.

The $\cn=4$ SYM theory includes four adjoint fermions $\psi_a$ and
six adjoint scalars $\phi^i$, all massless. If somehow these
fields would acquire masses, then at energies much lower than
these masses one would remain with pure Yang-Mills theory, which
is quite similar to QCD\footnote{This statement is only true for
small $\gym^2N$. At large $\gym^2N$, one finds that
$\Lambda_{\text{YM}}$ is of the same order of magnitude as the
masses of the fields we are trying to get rid of, thus, there is
no separation of scales in this case.}. One way to do this is to
add mass terms to all of these fields, i.e. adding to the SYM
lagrangian the term $\mathcal
O=M^2\tr(\phi^i\phi^i)+m_{ab}\tr(\bar\psi_a\psi_b) +\text{c.c.}$.
The problem with this is that the scalar part is not a chiral
operator, thus, this deformation cannot be studied in the
supergravity approximation. Alternatively, one may try to add the
chiral operator $(M^2)_{ij}\tr(\phi^i\phi^j)$ with $(M^2)_{ii}=0$,
but then some of the scalars would have negative mass-squares,
leading to instabilities. A way around this is to introduce masses
only for the fermions, breaking some (or all) of the
supersymmetry. In this paper we focus on introducing an equal mass
$m$ for all four fermions. Such a deformation breaks the
supersymmetry completely, hence we denote the theory one ends up
with as $\cn=0^*$ Yang-Mills theory. After the deformation, the
scalar mass term isn't protected, so one expects that the scalars
acquire mass quantum mechanically. From dimensional
considerations, the acquired mass squared must be proportional to
$m^2$. If it happens that the induced mass squared matrix is
strictly positive, then $\ave{\phi^i}=0$ will be a
(perturbatively) stable vacuum of the theory, and at the IR (well
below $m^2$) the theory becomes pure Yang-Mills with SU($N$) gauge
group. More generally, for this scenario to work, it is an
essential requirement that there is a stable vacuum at a value of
the scalars for which a non-Abelian gauge group remains unbroken.
Thus, this paper studies the viability of the $\cn=0^*$ theory as
a way to learn about pure YM theory (or theories similar to it).

In this paper we study the $\cn=0^*$ theory and compare the
results to those obtained on the supergravity side of the
correspondence. In section \ref{CFTsec} we study the scalar
effective potential and find that it has a perturbatively stable
vacuum. We also analyze the flow of the scalar mass-squared and
find some indications that it becomes positive at low energies,
implying that $\ave{\phi^i}=0$ is indeed a stable vacuum. In
section \ref{sugrasec} we review the analogue computation
preformed on the deformed AdS by brane probing the supergravity
background (the results described were obtained in
\cite{Babington:2002qt}), and compare it with the weak coupling
computation.
\section{The Deformed $\cn=4$ SYM}\label{CFTsec}
The \nef SYM lagrangian (with $\theta=0$) is given by
\begin{multline}\label{neflag}
    \lag=\frac1{\gym^2}\tr\bigg\{-\frac12F_{\mu\nu}F^{\mu\nu}
    -i\bar\psi^a\slhs{D}{5}\psi_a-D_\mu\phi^iD^\mu\phi^i
    +\\+
    \sum_{k=1}^3\tr\Big(C^k_{ab}\bar\psi_a[\phi^{2k-1},\psi_b]
    +iB^k_{ab}\bar\psi_a\gamma_5[\phi^{2k},\psi_b]\Big)
    +\frac12[\phi^i,\phi^j][\phi^i,\phi^j]\bigg\}
\end{multline}
where $\psi_a$ are the four Majorana fermions of the theory in
Dirac notation. The 4$\times$4 Yukawa coupling matrices $B^k_{ab}$
and $C^k_{ab}$ satisfy the algebra
\begin{align*}
    &[C^k,C^l]=2\eps^{klm}C^m,&
    &[B^k,B^l]=2\eps^{klm}B^m,&
    &[C^k,B^l]=0,\\
    &\{C^k,C^l\}=-2\delta^{kl},&
    &\{B^k,B^l\}=-2\delta^{kl}.
\end{align*}
The exact form of the $B$ and $C$ matrices we use may be read from
\eqref{masmat} below. The full SO(6)$_R$ is hidden in this
notation, since the fermions transform non trivially.

We deform the $\mathcal N=4$ SYM theory by adding the operator
\begin{equation}\label{deformbeta}
    \mathcal O_\beta=\frac{m}{2\gym^2}\sum_{a=1}^4\bar\psi_a\left(\cos\beta+i\gamma_5\sin\beta\right)\psi_a
\end{equation}
This deformation breaks the global symmetry
SU$(4)_R\ra$SO$(4)\simeq$SO$(3)\times$SO$(3)\times \Zf_2$. The six
scalars break into two groups of three that transform under the
SO$(3)\times$SO$(3)$ as $(3_v,1)$ and $(1,3_v)$. An SO(6)$_R$
transformation that rotates between the two triplets with an angle
$\delta\beta$ leaves all terms in the original lagrangian
invariant, but changes \eqref{deformbeta} by
$\beta\ra\beta+\delta\beta$.

In the following we calculate the one-loop corrections to the
scalar effective potential in a specific direction,
$\phi^i=\chi\d^{i,1}T_1$, where $T_1$ is the SU$(N)$ generator
\begin{equation*}
    T_1=\frac1{\sqrt{2N(N-1)}}\diag(1,1,\dotsc,1,1-N)
\end{equation*}
Notice that studying the scalar potential in a different direction
in the $\phi^1$-$\phi^2$ plane is equivalent to studying the
potential in the $\phi^1$ direction with a different angle $\beta$
in the deformation. On the supergravity side of the correspondence
the above choice describes putting $N-1$ branes at the origin and
probing with one brane away from the others. The supergravity side
is discussed in the next section.

\subsection{The Effective Potential}
In the supersymmetric theory ($m=0$) the effective potential is
flat in this specific direction to all orders in perturbation
theory. Thus, in order to find the 1-loop effective potential, we
calculate the contribution from diagrams with fermion loops:
\begin{fmffile}{1loop}
\begin{align*}
    V_\psi=\quad\bullet\quad+\quad\parbox{20mm}{\begin{fmfgraph*}(90,70)
    \fmfpen{thick}
    \fmfleft{in}
    \fmfright{out}
    \fmf{dashes,tension=2}{in,v1}
    \fmf{plain,left}{v1,v2,v1}
    \fmf{dashes,tension=2}{v2,out}
\end{fmfgraph*}}\quad\qquad+\qquad\parbox{20mm}{
\begin{fmfgraph*}(80,50)
    \fmfpen{thick}
    \fmfleft{in1,in2}
    \fmfright{out1,out2}
    \fmf{dashes,tension=1.8}{in1,vv1}
    \fmf{dashes,tension=1.8}{in2,vv2}
    \fmf{plain,left=0.4}{vv1,vv2,vv4,vv3,vv1}
    \fmf{dashes,tension=1.8}{vv3,out1}
    \fmf{dashes,tension=1.8}{vv4,out2}
\end{fmfgraph*}}\quad\qquad+\dotsb,
\end{align*}
\end{fmffile}
\noindent and then subtract the same result with $m=0$, in order
to take the bosonic loops contribution into account. Using a
procedure similar to that of Coleman and Weinberg
\cite{Coleman:1973jx}, the contribution from the massive fermions
at one-loop is
\begin{gather*}
    V_\psi(\chi,\beta)=-i\sum_{n=1}^\infty\frac{1}{2n}\tr(T_{1G}^{2n})\chi^{2n}(-1)^n\tr(C_1^{2n})
    \int\frac{d^4k}{(2\pi)^4}\tr\left\{\left[\frac{\c_\mu
    k^\mu+m(c_\beta-is_\beta\c^5)}{k^2-m^2}\right]^{2n}\right\}
\end{gather*}
With $c_\beta=\cos\beta\,,\,s_\beta=\sin\beta$, $T_{1G}$ is the
adjoint representation matrix of the generator $T_1$ and $C_1$ is
the Yukawa coupling matrix defined in \eqref{neflag}. They satisfy
\begin{align*}
    &\tr(C_1^{2n})=4(-1)^n
    &&\tr(T_{1G})^{2n}=\frac{N^n}{(2(N-1))^{n-1}}
    \approx 2^{1-n}N
\end{align*}
where we have taken the large $N$ limit\footnote{We are taking the
large $N$ limit since we are interested in comparing our results
to the string theory results which are valid at large $N$.
However, our one-loop computations in this section can be easily
performed also for finite $N$.}.

In order to calculate $\tr\big[\c_\mu
k^\mu+m(c_\beta-is_\beta\c^5)\big]^{2n}$, consider the general
term in the trinomial expansion: it has $s_1$ factors of
$mc_\beta$, $s_2$ factors of $\slh k$ and $s_3$ factors of
$-im\c^5s_\beta$. Let us denote the number of $\slh k$s to the
left most $\c^5$ as $l_1$, the number of $\slh k$s from the left
most $\c^5$ to the next $\c^5$ as $l_2$, and so on till
$l_{s_3+1}$, e.g.
\begin{equation*}
    \tr[\underbrace{\slh k\dotsm\slh k}_{l_1}\c^5
    \underbrace{\slh k\dotsm\slh k}_{l_2}\c^5
    \slh k\dotsm\underbrace{\dotsm\slh k}_{l_{s_3}}\c^5
    \underbrace{\slh k\dotsm\slh k}_{l_{s_3+1}}]=\begin{cases}
    (-1)^{l_2+l_4+\dotsb+l_{s_3}}4(k^2)^{s_2/2} & s_2,s_3\in2\Zf\\
    0&\text{otherwise}\end{cases}
\end{equation*}
Thus, the combinatorial problem may be defined as follows: put
$s_2=2q_2$ balls in $s_3+1=2q_3+1$ cells. Find the difference
between the number of configuration where the number of balls in
the first $q_3$ cells is even and the configuration where the same
quantity is odd. This number is
\begin{multline*}
    \sum_{p=0}^{2q_2}(-1)^p\binom{2q_2-p+q_3}{2q_2-p}\binom{p+q_3-1}{p}
    =\sum_{p=0}^{2q_2}(-1)^p\binom{-q_3-1}{2q_2-p}\binom{-q_3}{p}=\\=
    \sum_{p=0}^{q_2}(-1)^p\binom{-1}{2q_2-2p}\binom{-q_3}{p}=
     \sum_{p=0}^{q_2}(-1)^p\binom{-q_3}{p}=\binom{q_3+q_2}{q_2}
\end{multline*}
(we use generalized binomial coefficients). The second equality is
a formula taken from \cite{formula}. We find,
\begin{multline*}
    \tr\left[\c_\mu k^\mu+m(c_\beta-is_\beta\c^5)\right]^{2n}=\\=4\sum_{q_2=0}^n
    \sum_{q_3=0}^{n-q_2}\binom{2n}{2q_2+2q_3}\binom{q_2+q_3}{q_2}
    \big(c_\beta^2m^2\big)^{n-q_2}(k^2)^{q_2}(-)^{q_3}\tan^{2q_3}\beta
\end{multline*}
Thus,
\begin{multline*}
    V_\psi(\chi,\beta)=-16Ni\int\frac{d^4k}{(2\pi)^4}\times\\
    \times \sum_{n=1}^\infty\sum_{q_2=0}^n\sum_{q_3=0}^{n-q_2}\frac1n
    \binom{2n}{2q_2+2q_3}\binom{q_2+q_3}{q_2}
    \Big(\frac{\chi^2}{2}\Big)^n(-)^{q_3}\tan^{2q_3}\beta
    \frac{\big(k^2\big)^{q_2}\big(c_\beta^2m^2\big)^{n-q_2}}{(k^2-m^2)^{2n}}
\end{multline*}
This sum may be reorganized
\begin{equation*}
    \sum_{n=1}^\infty\sum_{q_2=0}^n\sum_{q_3=0}^{n-q_2}=
    \sum_{q_2,q_3=0}^\infty \sum_{n=\max(1,q_2+q_3)}^\infty=
    \sum_{n=1}^\infty\delta_{q_2,0}\delta_{q_3,0}
    +\sum_{q_3=1}^\infty \sum_{n=q_3}^\infty\delta_{q_2,0}
    +\sum_{q_2=1}^\infty\sum_{q_3=0}^\infty \sum_{n=q_2+q_3}^\infty
\end{equation*}
Respectively, these three terms yield
\begin{multline*}
    V_\psi(X,\beta)=-16iNm^4\int\frac{d^4q}{(2\pi)^4}
    \bigg\{
    -\log\Big[\frac{(q^2-1)^2-X^2\cos^2\beta}{(q^2-1)^2}\Big]-\\
    -\frac12\log\Big[\frac{(q^2-1)^4-2(q^2-1)^2X^2\cos2\beta+X^4}{((q^2-1)^2-X^2\cos^2\beta)^2}\Big]
    -\\\shoveright{-\frac12\log\Big[\frac{(q^2-1)^2(q^4+1+X^4-2q^2(1+X^2)-2X^2\cos2\beta)}
    {(q^2-1)^4-2(q^2-1)^2X^2\cos2\beta+X^4}\Big]
    \bigg\}=}\\=8iNm^4\int\frac{d^4q}{(2\pi)^4}\log\Big[\frac{q^4+1+X^4-2q^2(1+X^2)-2X^2\cos2\beta}{(q^2-1)^2}\Big]
\end{multline*}
where we have introduced the dimensionless quantities
\begin{align*}
    &X=\frac1{\sqrt2}\,\frac{\chi}{m}, &&q^\mu=\frac{k^\mu}m.
\end{align*}
As noted
above, adding the bosonic loops contribution to the effective
potential is the same as subtracting the fermionic contribution
with $m=0$, so the effective potential (up to counter-terms) is
\begin{multline}
    \vef=8iNm^4\int\frac{d^4q}{(2\pi)^4}\log\Big[\frac{q^4+1+X^4-2q^2(1+X^2)-2X^2\cos2\beta}{(q^2-1)^2}\Big]-\\
    -8iNm^4\int\frac{d^4q}{(2\pi)^4}\log\Big[\frac{(q^2-X^2)^2}{q^4}\Big]
\end{multline}
Rotating to Euclidean space ($q_0=-iq_4$) and introducing the 't
Hooft coupling $g_t^2=\gym^2N$,
\begin{equation}\label{vefct}
    \gym^2\vef=8g_t^2m^4\int\frac{d^4q}{(2\pi)^4}\log\Big[\frac{q^4\big(q^4+1+X^4+2q^2(1+X^2)-2X^2\cos2\beta\big)}{(q^2+X^2)^2(q^2+1)^2}\Big]
\end{equation}
In general, one should introduce counter-terms into an effective
potential for all relevant and marginal operators consistent with
the symmetry, this is how renormalization is reflected in the
Coleman-Weinberg formalism. The result we got in \eqref{vefct} has
only a $\Lambda^2$ divergence (as a function of a cutoff
$\Lambda$), the $\log\Lambda^2$ vanishes since the potential must
vanish in the original theory ($m=0$). Thus, we only need to add a
mass counter-term, which we choose such that
\begin{equation}\label{rencon}
    M_1^2\cos^2\beta+M_2^2\sin^2\beta=\frac1{\gym^2}\diff{^2V}{\chi^2}\bigg\vert_{\chi=0}
\end{equation}
We have
\begin{multline*}
    \frac{\gym^{2}\vef}{m^4}=\frac{M_1^2\cos^2\beta+M_2^2\sin^2\beta}{m^2}X^2
    +\\+\frac{g_t^2}{\pi^2}\bigg\{\int
    dq\,q^3\log\Big[\frac{q^4(q^4+1+X^4+2q^2(1+X^2)-2X^2\cos2\beta)}{(q^2+1)^2(q^2+X^2)^2}\Big]
    +2(2+\cos2\beta)X^2\log\Lambda-X^2\bigg\}
\end{multline*}
Performing the integration we find
\begin{multline}\label{poten}
    \frac{\gym^{2}\vef}{m^4}=\frac{M_1^2\cos^2\beta+M_2^2\sin^2\beta}{m^2}X^2+\frac{g_t^2}{4\pi^2}\bigg\{-2X^2-12X^2\cos^2\beta
    -X^4\log X^4+\\+4X(1+X^2)\cos\beta\log\frac{1+2X\cos\beta+X^2}{1-2X\cos\beta+X^2}\\
    +(1+2X^2+4X^2\cos^2\beta+X^4)\log\big(1-2X^2\cos2\beta+X^4\big)\bigg\}
\end{multline}
So far we have been treating the angle $\beta$ as a parameter of
the deformation \eqref{deformbeta}. However, by a SO(6) rotation,
the calculation preformed above with a specific $\beta$ is
equivalent to deforming the \nef lagrangian by a standard mass
term (putting $\beta=0$ in \eqref{deformbeta}) and studying the
potential in a direction $\phi^1+i\phi^2=\sqrt2mXe^{i\beta}$. Thus
\eqref{poten} is the 1-loop effective potential in the
$(\phi_1,\phi_2)$ plane (which is related by the remaining
R-symmetry to the other scalar directions).

Let us analyze the effective potential we obtained without
considering its range of validity. For large $X$,
$\gym^2\vef\approx g_t^2m^4X^2\log X^4$, i.e. this flat direction
is lifted to a stable potential at one-loop. For small $X$,
\begin{equation}\label{smallX}
   \frac{\gym^2\vef}{m^4}=\frac{M_1^2\cos^2\beta+M_2^2\sin^2\beta}{m^2}X^2-\frac{g_t^2}{4\pi^2}X^4\log
   X^4+O(X^4)
\end{equation}
For $M_1^2,M_2^2\geq0$ the effective potential has a minimum at
$X=0$, this is a stable vacuum. If $M_i^2<0$ for  $i=1$ or $2$ (or
both), $X=0$ becomes a saddle point (maximum) and we find a
minimum for some $X>0$. In the latter case both the gauge and the
R-symmetry groups are broken (notice that even if $M_1^2=M_2^2$
the potential isn't radially symmetric although the expansion
\eqref{smallX} is, i.e. R-symmetry is completely broken).

There are two types of logarithms in this effective potential. The
first, $\log X$, diverges at $X=0$, and the other logarithms
diverge at the four points $X=\pm e^{\pm i\beta}$ (notice that $X$
is defined to be real, so these points are not physical unless
$\beta=0$). The $X=0$ divergence is well expected since the
$\cn=0^*$ theory is IR divergent. The other divergences are easy
to see by using Weyl notation for the fermions. The terms in the
lagrangian that are quadratic in the fermions may be written
$im\tr(A_{ab}\lambda_a^T\sigma^2\lambda_b)+$c.c. (the $\tr$ is
over the gauge group), where
\begin{equation}\label{masmat}
    A=\gym^{-2}\begin{pmatrix}
      e^{i\beta} & X^1+iX^2 & X^3+iX^4 & -X^5+iX^6 \\
      X^1+iX^2 & e^{i\beta} & X^5+iX^6 & X^3-iX^4 \\
      X^3+iX^4 & X^5+iX^6 & e^{i\beta} & X^1-iX^2 \\
      -X^5+iX^6 & X^3-iX^4 & X^1-iX^2 & e^{i\beta} \\
    \end{pmatrix}
\end{equation}
and $X^i=\phi^i/m$. By turning off all the scalars except $X^1$
and $X^2$, and treating $X^1+iX^2$ as a complex scalar, we see
that the points $(X^1+iX^2)=\pm\ e^{\pm i\beta}$ are exactly the
values where \eqref{masmat} has a zero eigenvalue (exactly one
eigenvalue vanishes at each root). Therefore, the divergences of
the effective potential at these values correspond to values of
the VEV where one fermion becomes massless.

Higher loop corrections to this potential would be polynomials in
$g_t^2\log X$ and $g_t^2\log(X-r_i)$ for $r_i=\pm e^{\pm i\beta}$.
At constant $X\neq0$ equation \eqref{poten} would be a good
approximation to the full potential as long as $g_t$ is small
enough. Thus, although the potential near $X=0$ is not
trustworthy, the existence and location (or lack of existence) of
stable vacua for $X>0$ are reliable for small enough $g_t$.
\subsection{Renormalization Group Flow}\label{rgfssc}
In the above we saw that the location of the stable vacuum of the
theory depends on the sign of the arbitrary mass renormalization
parameters $M_1^2$ and $M_2^2$, introduced in \eqref{rencon}. At
the UV the $\cn=0^*$ theory is expected to flow to the $\cn=4$
theory, hence, it makes sense to define the scalar masses counter
terms such that the scalar masses vanish at the UV and then follow
its sign as we flow to the IR. To do this we follow the standard
procedure of renormalization (in momentum space) in order to
analyze the flow of $M_i^2$ to lowest order.

We calculate the scalar two-point function in the same direction
($\phi^i=\chi\d^{i,1}T_1$) as before. The diagram with massive
fermions is
\begin{fmffile}{ren}
\begin{align*}
    \Pi(m)&=\quad\parbox{20mm}{\begin{fmfgraph*}(90,70)
    \fmfpen{thick}
    \fmfleft{in}
    \fmfright{out}
    \fmf{dashes,tension=2}{in,v1}
    \fmf{plain,left}{v1,v2,v1}
    \fmf{dashes,tension=2}{v2,out}
\end{fmfgraph*}}\quad\qquad=\\&=\tr[(C_1)^2]f_{1ml}f_{1ml}\tr\int\frac{d^dk}{(2\pi)^d}\frac{k_\mu\gamma^\mu+
    m(c_\beta-is_\beta\gamma_5)}{k^2-m^2+i\eps}\frac{(p+k)_\nu\gamma^\nu+
    m(c_\beta-is_\beta\gamma_5)}{(p+k)^2-m^2+i\eps}=\\&
    =-\frac{ig_t^2}{\pi^2\gym^2}\cdot\frac{\hat p^2+m^2(4+2\cos2\beta)}{\eps}+O(\eps^0)
\end{align*}
\end{fmffile}
where $\hat p$ is the Wick rotated momentum and we have employed
dimensional regularization $(\eps=4-d)$. The singular part of the
scalar two-point function vanishes in the original $\cn=4$ theory.
Thus, to get the full scalar 2-point function (including the
bosonic contributions), we subtract the same with $m=0$,
\begin{equation*}
    \Pi_{XX}=\Pi(m)-\Pi(0)=-\frac{ig_t^2}{\pi^2\gym^2}\cdot\frac{m^2(4+2\cos2\beta)}{\eps}+O(\eps^0)
\end{equation*}
From this counter term, one finds the RG equations for the scalar
masses,
\begin{align*}
    &\mu\pdf{M_1^2}\mu=-\frac{6g_t^2m^2}{\pi^2}+o(g_t^4)&
    &\mu\pdf{M_2^2}\mu=-\frac{2g_t^2m^2}{\pi^2}+o(g_t^4)\\
    \Rightarrow\quad
    &M_1^2=-\frac{6g_t^2m^2}{\pi^2}\log(\mu/\mu_0)+o(g_t^4)&
    &M_2^2=-\frac{2g_t^2m^2}{\pi^2}\log(\mu/\mu_0)+o(g_t^4)
\end{align*}
Thus, if one sets the parameters at the cutoff $\mu_0$ to be such
that the theory at $\mu_0$ will be $\mathcal N=4$ SYM deformed by
the mass term \eqref{deformbeta}, then the theory will flow such
that the scalars get a positive mass-squared. This remains valid
as long as $g_t^2\log(\mu/\mu_0)$ remains small. It is then
reasonable to assume that the free parameters, $M_i^2$, introduced
in \eqref{poten} should be positive, implying that $X=0$ is a
stable vacuum.
\section{The Scalar Potential from AdS/CFT Correspondence}\label{sugrasec}
We briefly review the formalism used in \cite{Babington:2002qt} to
calculate the scalar potential from the AdS side of the
correspondence.
\subsection{Deformed AdS}\label{defadsss}
The AdS/CFT correspondence maps between $\N=4$ Super Yang-Mills
deformed by a chiral operator and type IIB Superstrings on a
modified \ads string theory background, which can be approximated
by supergravity at large 't Hooft coupling. Deforming the CFT by a
scalar (and chiral) operator $\O$ with scaling dimension $\Delta$
is dual to turning on a scalar field in the \ads with mass:
\begin{equation}
    m^2=\Delta(\Delta-4)
\end{equation}
A consistent and relatively easy process to find the deformed
supergravity background is to find a corresponding AdS$_5$ $\N=8$
gauged supergravity background and rely on the consistent
truncation conjecture to lift the solutions to the full 10D Type
IIB Supergravity. The scalar field corresponding to the \nez
deformation \eqref{deformbeta} was identified in
\cite{Girardello:1999bd} as the scalar generated by the lowest
spherical harmonic of the $\mathbf{10_c}$ representation of the
global $\SO(6)$. The relevant equation of motion can be derived
from the gauged supergravity action:
\begin{equation}
    S=\int d^5x\sqrt
    g\left(R+\frac12\prt_\mu\l\prt^\mu\l+V(\l)\right),
\end{equation}
\begin{equation}
    V(\l)=-\frac32\left(1+\cosh^2\l\right).
\end{equation}
We assume the following ansatz for the 5-d metric:
\begin{equation}
    ds^2=e^{2A(r)}dx^\mu dx_\mu+dr^2\hsp2 \mu=0,1\ldots 4
\end{equation}
Noticing that $\l$ must depend solely on $r$, the equation of
motion reduces to:
\begin{gather}\label{eom1}
    \l''+4A'\l'=\pdf{V}{\l}=-2\cosh\l\sinh\l\\\label{eom2}
    6A'^{\,2}=\l'^{\,2}-2V(\l)=\l'^{\,2}+3\left(1+\cosh^2\l\right)
\end{gather}
At large $r$ the solution must be asymptotically AdS$_5$, e.g.
$A(r)\xrightarrow{r\rightarrow\infty}r$ and
$\l(r)\xrightarrow{r\rightarrow\infty}0$. The equation of motion
can be solved to first order at this limit to find the asymptotic
behavior:
\begin{gather}\label{mkpar}
    V(\l)\approx-3-\frac32\l^2+0(\l^4)\rar2
    \l(r\rightarrow\infty)=\mathcal Me^{-r}+\mathcal Ke^{-3r}
\end{gather}
The exact equation of motion (\ref{eom1}),(\ref{eom2}) can be
solved numerically (a full discussion of the solutions for
different boundary conditions $(\mathcal M,\mathcal K)$ is given
in \cite{Babington:2002qt}). The parameters $\mathcal M$ and
$\mathcal K$ correspond respectively to the coefficient of the
deformation \eqref{deformbeta} in the CFT and the VEV of this
operator. It was shown in \cite{Babington:2002qt} that the
numerical solutions of the equations of motion are singular at
finite $r$ for all values of $(\mathcal M,\mathcal K)$. Thus,
supergravity breaks down near this point and one should really
find a full string background which should be non-singular (as was
suggested in \cite{Polchinski:2000uf} for the $\cn=1^*$ theory).
Still, one expects that far from the singularity the full
background will be similar to the supergravity background, so
supergravity should be a good approximation for the computations
performed there. The parameter $\mathcal K$ is determined in
principle by the behavior of the solution for small $r$, but since
the solution is singular, it remains undetermined in the
supergravity approximation.
\subsection{Brane Probe Potential}
The scalar potential in the strongly coupled \nez theory can be
calculated in the deformed \ads background by the method of brane
probing. Using the Born-Infeld action for a D3-brane probe
separated by a distance $r$ from the center of the AdS$_5\times
S^5$, it is easy to find the induced potential on the radial
coordinate of the probe location. The radial coordinate of the
probe is mapped to the location of the VEV $X$, discussed in
section \ref{CFTsec}. The exact lifting of the solution to 10D and
the probe potential calculation was done in
\cite{Babington:2002qt}. The result is
\begin{equation}\label{pro_pot1}
    V_{\text{probe}}(\chi)=\tau_3e^{4A(r)}\left[\xi(r)-\frac{dA}{dr}\right],
\end{equation}
where $A(r)\,,\,\l(r)$ are the solutions to
(\ref{eom1}),(\ref{eom2}) and $\tau_3$ is the D3-brane tension.
The 10D deformed AdS metric is divided to a part tangent to the
D3-brane probe ($ds^2_{1,4}$) and a part to it
\begin{equation*}\label{pro_met1}
    ds_{10}^2=\xi^{1/2}ds^2_{1,4}+\xi^{-\frac32}\left[\cos^2\alpha\,\xi_-d\Omega_+^2
    +\sin^2\alpha\,\xi_+d\Omega_-^2+\xi^2d\alpha^2\right]
\end{equation*}
\begin{align*}\label{pro_pot2}
    &\xi_\pm=\cosh^2\l\pm\cos^2(2\alpha)\sinh^2\l\cr
    &\xi^2=\xi_+\xi_-=\cosh^4\l-\cos^2(2\alpha)\sinh^4\l
\end{align*}
In the above, the 6-dimensional space perpendicular to the
D3-brane probe is parameterized in terms of two S$^2$ spheres, a
radial coordinate $r$ and an angle $\alpha$. The two S$^2$ spheres
are realized by the two constraints
\begin{gather*}
    \sum_{i=1}^3 \left(U_+^i\right)^2=r^2\cos\alpha\hsp1,\hsp1 \sum_{i=1}^3 \left(U_-^i\right)^2=r^2\sin\alpha
\end{gather*}
Where the coordinates $U^i_+,U^i_-$ maps to $\phi^1,\phi^3,\phi^5$
and $\phi^2,\phi^4,\phi^6$  of section \ref{CFTsec}, up to
re-parametrization of the radial coordinate $r$. This
re-parametrization between the radial coordinate $r$ (the distance
of the D3-brane from the center of the AdS$_5$) and the VEV $X$
(which is used in section \ref{CFTsec}) is found by redefining the
$r$ field in the Born-Infeld action such that it is normalized
exactly as X in section \ref{CFTsec}. This re-parametrization is
given by
\begin{equation}\label{repar}
    X(r)=\frac{1}{2\pi\sqrt{2m^2\alpha'}}\int_0^re^{A(r')}\xi^{1/2}(r')dr'
\end{equation}
The probe potential is maximal at $\alpha=\pi/4$ and has a period
of $\pi$. Comparing to the behavior in $\beta$ of the \nez
effective potential we find the identification
$\alpha=\beta+\pi/4$.

A numerical computation produces the probe potential shown in
Figure 1 (right side). The numerical computation fails at a small
value of $\chi_0$, and the graph is produced by cutting the
potential for $\chi<\chi_0$. Due to the numerical difficulty the
translation between $\chi$ and $r$ is defined up to addition of a
small constant value (which cannot be computed using this
approach).
\begin{figure}[t]
\hspace{-1cm}%
\includegraphics[height=7.0cm]{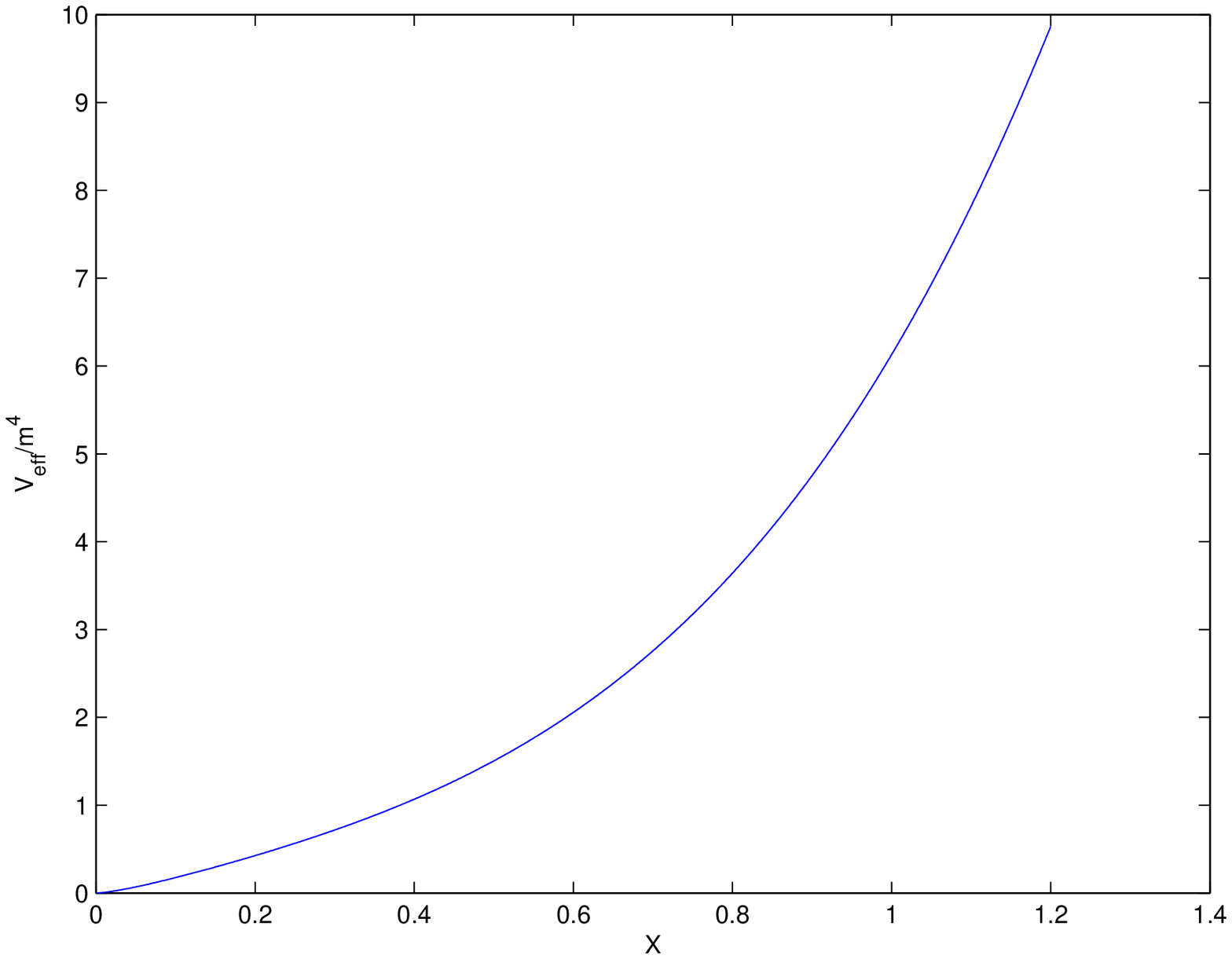}%
\hspace{0.5cm}%
\includegraphics[height=6.7cm]{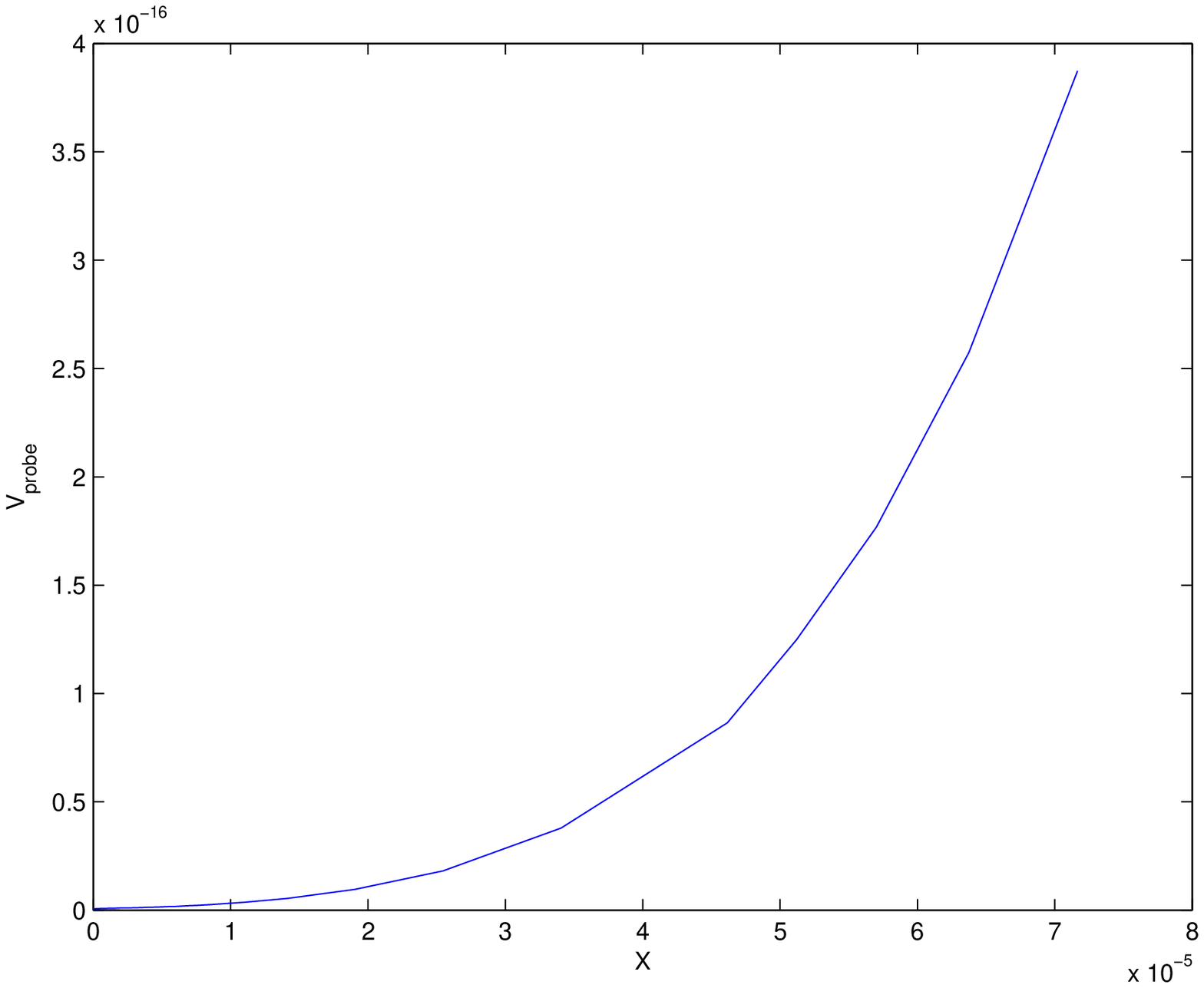}
\caption{The scalar effective potential in the deformed CFT with
$M=0$ and $\beta=-\pi/4$ (left) and the probe potential in the
deformed AdS with $\alpha=0$ (right)}
\end{figure}
The solution of the equations of motion depends on boundary
conditions related to the asymptotic behavior of the field $\l$.
The numerical analysis was done for $(\mathcal M,\mathcal
K)=(1,0)$. As discussed at the end of \ref{defadsss}, the value of
$\mathcal K$ remains undetermined in the supergravity analysis
(one expects $\mathcal K$, which corresponds to the gluino
condensate in the CFT, to be of order unity in units of $m^3$).
Fortunately, the qualitative behavior of the potential does not
seem to depend much on this number. Allowing other values for the
parameters ($\mathcal K>0$) does not change the qualitative
features shown in figure 1.
\section{Conclusions}
We have calculated the effective potential of the \nez theory and
shown that 1-loop corrections make the potential stable in
specific directions that are flat at tree level (i.e. flat in the
unmodified \nef SYM theory). Note that although the gauge symmetry
and the global R-symmetry restrict the general form of the
potential, they do not fix it completely and there remain
unexplored directions which were flat in \nef SYM.

It is interesting that the scalar potential we found in the \nez
theory is qualitatively similar to the probe potential in its
supergravity dual, although their ranges of validity do not
coincide.

We have seen some indications that the vacuum of the \nez theory
is at $\ave{\phi^i}=0$, implying that both the gauge symmetry and
the R-symmetry remain unbroken. The location of the vacuum depends
on the sign of the parameters $M_1^2$ and $M_2^2$ introduced in
\eqref{poten}. In subsection \ref{rgfssc} we gave arguments why
these parameters should be positive, but they are not a proof,
since they fail at low energies (where the theory becomes strongly
coupled and the perturbative description fails). The brane probe
potential in the supergravity approximation also indicates the
same conclusion (a stable symmetric vacuum). However, it too fails
in the interior of the AdS, implying the approximation breaks down
and should not be trusted for small $X$.

The failure of the supergravity approximation in the interior of
the AdS is a hint for stringy physics in this area. The true
string vacuum dual to \nez is likely to be described by some
extended brane configuration, analogous to the configurations
found by Polchinski and Strassler \cite{Polchinski:2000uf} for the
string dual of the \neo theory. Among its other advantages,
knowing the full string background should allow one to calculate
the value of the parameter $\mathcal K$ (introduced in
\eqref{mkpar}), thus picking the right solution asymptotically for
a given $\mathcal M$. It is also interesting to see if the
qualitative similarity between the effective potential in the \nez
theory and the probe potential in its supergravity dual, as
depicted in Figure 1, will abide outside the supergravity
approximation.

\acknowledgments We would like to thank our advisors Ofer Aharony
and Micha Berkooz for their guidance and support throughout this
project. We would also like to thank Jacques Distler for his
illuminating remarks that were communicated to us through Ofer
Aharony. This work was supported in part by the Israel-U.S.
Binational Science Foundation, by the ISF Centers of Excellence
Program, by the European RTN network HPRN-CT-2000-00122, and by
the Minerva foundation.
\bibliographystyle{utphys}
\bibliography{0402215_v2}
\end{document}